\documentclass[preprint showpacs,preprintnumbers,amsmath,amssymb]{revtex4}
\usepackage{longtable}
\usepackage{xcolor}
\usepackage{graphicx}
\usepackage{dcolumn}
\usepackage{bm}
\usepackage{multirow}
\usepackage{natbib}
\usepackage{color}
\usepackage{cases} 

\begin{document}
\preprint{APS/123-QED}

\title{Reactive collisions between electrons and  BeT$^+$: Complete set of thermal rate coefficients up to 5000 K.}

\author{N. Pop$^{1}$}
\author{F. Iacob$^{2}$}
\email[]{felix.iacob@e-uvt.ro}
\author{S. Niyonzima$^{3}$}
\author{A. Abdoulanziz$^{4}$}
\author{V. Laporta$^{5}$}
\author{D. Reiter$^{6}$}
\author{I. F. Schneider$^{4,7}$}
\author{J. Zs Mezei$^{8,4}$}
\affiliation{$^{1}$Department of Physical Foundations of Engineering, Politehnica University of Timisoara, 300223 Timisoara, Romania}%
\affiliation{$^{2}$Department of Physics, West University of Timisoara, 300223 Timisoara, Romania}%
\affiliation{$^{3}$D\'epartement de Physique, Fac. des Sciences, Universit\'e du Burundi, B.P. 2700 Bujumbura, Burundi}%
\affiliation{$^{4}$LOMC CNRS-UMR6294, Universit\'e le Havre Normandie, F-76058 Le Havre, France}%
\affiliation{$^{5}$Istituto per la Scienza e Tecnologia dei Plasmi, CNR, 70126 Bari, Italy}%
\affiliation{$^{6}$Institute for Laser and Plasma Physics, Heinrich-Heine-University, D-40225 D\"usseldorf, Germany}%
\affiliation{$^{7}$LAC CNRS-FRE2038, Universit\'e Paris-Saclay, F-91405 Orsay, France}%
\affiliation{$^{8}$Institute for Nuclear Research (ATOMKI), H-4001 Debrecen, Hungary}%
\date{\today}

\begin{abstract}
Rate coefficients for the dissociative recombination, vibrational excitation and vibrational de-excitation of the BeT$^{+}$ ion for all vibrational levels of its ground electronic state ($ X\ensuremath{^{1}\Sigma^{+}},v_{i}^{+}=0,\dots,27$) are reported,
including in the calculation the contribution of  super-excited states of the BeT complex pertaining to three electronic symmetries - $^{2}\Pi$, $^{2}\Sigma^{+}$,  and  $^{2}\Delta$.
These data are suitable for the kinetic modeling of beryllium and tritium containing plasma,
as encountered in magnetic fusion devices with beryllium walls (JET, ITER). In the present study we restrict ourselves to incident electron energies from 10$^{-3}$ up to $2.7$ eV, and to electron temperatures between $100$ and $5000$ K, respectively. Together with our earlier and closely related studies on the BeH$^{+}$ and BeD$^{+}$ systems,
this present work completes the isotopic coverage for the beryllium monohydride ions.
The vibrational energy (rather than the vibrational quantum state) is identified as a proper isotopic similarity parameter, e.g., for reduced but still isotopically correct plasma chemistry models.
\end{abstract}


\maketitle

\section{\label{sec:intro}Introduction}

The plasma in the peripheral region (plasma boundary) of the ITER fusion reactor~\cite{Kleyn2006} necessarily interacts with the first wall materials, which are mostly covered with the low Z metal beryllium in the main chamber. In the outer (near wall) plasma domains this leads to a complex diversity of interconnected physical and chemical elementary processes~\cite{Reiter-Springer2005}.
The fuel used in the activated phases in next step magnetic fusion reactors consists of deuterium and the tritium at equal concentrations. The d-t fusion reaction leads to helium and neutrons as products, which carry the excess energy from these fusion processes as kinetic energy. While the neutrons leave the plasma flame, the helium ash is confined and therefore magnetically guided out (together with the other plasma constituents) from the main plasma chamber into the so called divertor\cite{Kleyn2006,Reiter-Springer2005}. The latter is  equipped with pumps and high heat flux components (currently envisaged there: tungsten target surfaces).

Beryllium and tungsten are therefore considered as major plasma-facing materials for the ITER reactor~\cite{Reiter2012,Motojima2015,Mitteau2017} which, once released, act as plasma impurities. Thus it is of great importance to quantitatively understand their release, the transport, and the chemical reactivity of these metals in the plasma.
The main chamber beryllium wall is exposed to plasma heat and particle bombardment from the outer boundary plasma region. Identification and quantification of the various release candidate mechanisms, such as physical 
, chemical sputtering or other such processes, is still an active field of ongoing research in magnetic confinement fusion \cite{Reiter2012,Pamela2007,Borodin2016}. Chemically assisted physical sputtering contributes to the formation of beryllium atoms at surfaces; release mechanisms of molecular forms BeH or BeH$_2$ may also play a role. The beryllium atoms or beryllium hydrides in turn will enter into the plasma, fragment (dissociate) and radiate there, and/or form further molecular species like BeH, BeD and BeT (or their ions) in chemical reactions with the atoms of the fuel (H, D, T) or their molecules.

Besides the experimentally confirmed presence of Be containing neutral atomic and molecular species in magnetic fusion plasma, their ionic counterparts ($\rm Be^+$, $\rm BeX^+$, where X stands for H, D, or T) were also presumed early~\cite{Duxbury1998,Darby2018,Niyonzima2017} and meanwhile also been experimentally confirmed by spectroscopic methods~\cite{Brezinsek2014, Brezinsek2015, Krieger2013, Nishijima2008, Doerner2009}.
Even though the fuel hydrogen isotopes remain, by far, the dominant gas and charged components, in the colder edge/divertor regions collisions between comparatively low energy (0.1 -- 100 eV) electrons and BeX$^{+}$ ions can play a significant role for the overall transport and fragmentation pathways
of these impurities and hence, globally, for the fusion plasma flame purity.
Typically, in current fusion devices related impurity wall release rates are inferred only rather indirectly,
via intensity of light emission from the Be and BeX containing plasma constituents. However, this experimental procedure combines two unknowns: firstly the surface release mechanisms and rates, and, secondly, the (mostly) electron collision driven volumetric collisional processes (incl. fragmentation, excitation, etc.). Our present computational work on the BeX$^{+}$ family of ions contributes to a wider effort to separate these two unknowns, by isolating the latter with detailed theoretical cross section calculations, hence to make the former then better experimentally accessible.

The most relevant reactive collisions starting from BeX$^{+}$ ions that can take place are the following:
\begin{equation}\label{reac:dr}
\mbox{BeX}^{+}(v_{i}^{+}) +e^{-} \longrightarrow  \mbox{Be} + \mbox{X}\qquad\mbox{Dissociative Recombination (DR)}\qquad
\end{equation}
\begin{equation}\label{reac:ve}
\mbox{BeX}^{+}(v_{i}^{+}) +e^{-} \overset{v_i^+<v_f^+}{\longrightarrow}  \mbox{BeX}^{+}(v_{f}^{+}) +e^{-}\quad\mbox{Vibrational Excitation (VE)}\quad
\end{equation}
\begin{equation}\label{reac:vde}
\mbox{BeX}^{+}(v_{i}^{+}) +e^{-} \overset{v_i^+>v_f^+} {\longrightarrow} \mbox{BeX}^{+}(v_{f}^{+}) +e^{-}\enspace\mbox{Vibrational deExcitation (VdE)}\enspace
\end{equation}
\noindent and, at high collision energies - which will be a range addressed in future work:
\begin{equation}\label{reac:de}
\mbox{BeX}^{+}(v_{i}^{+}) +e^{-} \longrightarrow  \mbox{Be}^+ + \mbox{X}+e^-\qquad\mbox{Dissociative Excitation (DE)}\qquad
\end{equation}
\noindent where $v_i^+$ and $v_f^+$ denote the initial and final vibrational quantum numbers of the target cation.

In the framework of the Multichannel Quantum Defect Theory (MQDT)~\cite{Giusti1980,Seaton1983,Schneider1991,Jungen2011,Niyonzima2013,Mezei2019} we have performed nuclear dynamics calculations using previously computed molecular data~\cite{Niyonzima2013,Roos2009} updated for the heaviest beryllium monohydride isotopologues.
The present work continues and ends a series of studies we performed on the DR, VE and VdE - eqs.~(\ref{reac:dr}-\ref{reac:vde}) - 
of BeH$^+$~\cite{Niyonzima2017,Niyonzima2013,Laporta2017} and BeD$^+$~\cite{Niyonzima2018, Pop2019, Pop2017} with electrons of low to moderate collision energy.
Here we first present a complete study on the vibrational dependence of the DR, VdE and VE rate coefficients of BeT$^+$ cation.
Having thus completed the set of monohydride BeX$^{+}$ ion isotopologues enables us now to also derive an isotopic scaling law from this data-set (see Section 3), which greatly facilitates the use of this detailed information inside the physically and numerically already demanding
multi-physics multi-scale fusion boundary plasma computational tools, such as the current ITER boundary plasma suite of codes SOLPS-ITER \cite{Pitts2019} .

The paper is organized as follows: The introduction is followed by a brief description of the employed theoretical approach and some computational details. 
Section \ref{sec:res} presents our calculated rate coefficients. The paper ends with conclusions.

\section{Theoretical approach of the dynamics}\label{sec:theory}
In the present study, we used an MQDT-based method to study the electron-impact collision processes given by Eqs.~(\ref{reac:dr}--\ref{reac:vde}).
Along with other methods we use \cite{fi}, MQDT has proven to be a powerful and successful method when applied to several diatomic systems like H$_2^+$~\cite{Motapon2014}, N$_2^+$~\cite{Little2014}, CO$^+$~\cite{Moulane2018}. This approach, combined with the R-matrix technique, was also used to model satisfactorily - although less accurately - electronic collisions with poly-atomic ions like BF$_2^+$~\cite{Kokoouline2018} and NH$_2$CH$_2$O$^+$~\cite{Ayouz2019}.

The processes studied in the present paper involve ionization channels, describing the scattering of an electron on the molecular cation, and dissociation channels, accounting for atom-atom scattering. The mixing of these channels results in quantum interference of the {\it direct} mechanism - in which the capture takes place into a doubly excited dissociative state of the neutral system - and the {\it indirect} one -  in which the capture occurs \textit{via} a Rydberg state of the molecule, predissociated by the dissociative state.
 The direct mechanism dominates the reactive collisions in the cases of favorable crossings (in accordance with the Franck-Condon principle) between the potential energy curves of the dissociative states and that of the target ion. 
In both mechanisms the autoionization is in competition with the predissociation, and leads to the excitation/de-excitation of the cation. 
In the present calculations we account for vibrational structure 
for the target ion's ground electronic state and for the neutral's relevant electronic states, and neglect the rotational effects.
A detailed description of our theoretical approach has been given in our previous articles
- see for example ~\cite{Niyonzima2013}  and  ~\cite{Mezei2019} and references therein.
Here the general ideas and major steps are recalled only. 
Within a quasi-diabatic representation of the molecular states, and for a given set of conserved quantum numbers  {{of the neutral system - $\Lambda$ (projection of the electronic angular momentum on the internuclear axis), $S$ (total electronic spin) 
- the interaction matrix is built based on the couplings between
\textit{ionisation} channels -
	associated to
	the vibrational levels $v^+$ of the cation
	and
	to the orbital quantum number $l$ of the incident/Rydberg electron -
and
\textit{dissociation} channels $d_j$.
By adopting the second-order perturbative solution
for the Lippman-Schwinger integral equation \cite{Ngassam2003}, we compute the reaction matrix of our collision system in the reaction zone.

 A frame transformation is performed eventually, from the Born-Oppenheimer (short range) representation, characterized by $v$ and $\Lambda$ quantum numbers, valid for small electron-ion and nucleus-nucleus distances, to the close-coupling (long-range) representation, characterized by  $v^+,\Lambda^+$ (for the ion) and $l$ (orbital quantum number of the incident/Rydberg electron), valid for both large distances.
 This frame transformation relies on the quantum defects $\mu_{l}^{\Lambda}(R)$ describing the relevant Rydberg series built on the ionic core, and on the eigenvectors and eigenvalues of the K-matrix.

Based on the frame-transformation coefficients, we then build the physical scattering matrix organized in blocks associated to energetically {\it open} and/or {\it closed} ($O$ and/or $C$ respectively) channels: 
\begin{equation}\label{eq:solve3}
\boldsymbol{S}=\boldsymbol{X_{OO}}-\boldsymbol{X_{OC}}\frac{1}{\boldsymbol{X_{CC}}-\exp(-i2\pi\boldsymbol{ \nu})}\boldsymbol{X_{CO}}
.
\end{equation}

The first term in Eq.~({\ref{eq:solve3}}) is restricted to the open channels, resulting in the {\it direct} mechanism, and the second takes into account their mixing with the closed ones, resulting in the {\it total}, {\it i.e.} direct and indirect mechanism,  the denominator being responsible for the resonant patterns in the shape of the cross section \cite{Seaton1983}. Here the matrix $\exp(-i2\pi\boldsymbol{ \nu})$ is diagonal and relies on the effective quantum numbers $\nu_{v^{+}}$ associated to the vibrational thresholds of the closed ionisation channels.

Once we have the scattering matrix the {\it computation of the cross-sections} is straightforward. For a given target cation on vibrational level $v_i^+$ and for a given energy of the incident electron $\varepsilon$, the dissociative recombination and the vibrational transition - elastic scattering, excitation, de-excitation - cross sections are computed using, respectively:

\begin{equation}\label{drxsec1}
	\sigma_{diss \leftarrow {v^+_{i}}} = \frac{\pi}{4\varepsilon} \sum_{{l,\Lambda}}
	\rho^{{\Lambda}} \sum_{j} \left|S^{{\Lambda}}_{d_j, {l}v^+_{i}}\right|^2,
\end{equation}

\begin{equation}\label{vexsec1}
	\sigma_{{v^+_{f}} \leftarrow  {v^+_{i}}} = \frac{\pi}{4\varepsilon}\sum_{{l,l'\Lambda}}
\rho^{{\Lambda}} \left|S^{{\Lambda}}_{ {l'}{v^+_{f}}, {l}{v^+_{i}} } -
{\delta_{l'l}} \delta_{{v^+_{f}}{v^+_{i}}}\right|^2.
\end{equation}

\noindent
Here $\rho^{{\Lambda}}$ is the ratio between the spin and angular momentum multiplicities of the neutral and the target ion.
\begin{table}[t]
\caption{BeT$^+$ vibrational levels relative to the $v_i^+=0$. The values of dissociation energies are $D_e=2.795$ eV and $D_0=2.704$ eV.}
\label{table:1}
\begin{center}
\begin{tabular}{lcrclclc}
\hline
$v_i^+$ & $E_{v_i^+}$ (eV) & $v_i^+$ & $E_{v_i^+}$ (eV) & $v_i^+$ & $E_{v_i^+}$ (eV) & $v_i^+$ & $E_{v_i^+}$ (eV)\\
\hline
0 & 0.000 & 7 & 1.078 & 14 & 1.921 & 21 & 2.477\\
1 & 0.167 & 8 & 1.214 & 15 & 2.017 & 22 & 2.532\\
2 & 0.327 & 9 & 1.347 & 16 & 2.107 & 23 & 2.578\\
3 & 0.486 &  10 & 1.475 & 17 & 2.193 & 24 & 2.617\\
4 & 0.641 &  11 & 1.597 & 18 & 2.273 & 25 & 2.648\\
5 & 0.792 &  12 & 1.712 & 19 & 2.348 & 26 & 2.674\\
6 & 0.937 &  13 & 1.819 & 20 & 2.416 & 27 & 2.693\\
\hline
\end{tabular}
\end{center}
\end{table}

\section{Results and discussions}\label{sec:res}

Neglecting the very slight effects related to the change of the nuclear mass when Tritium replace Hydrogen, we have used the molecular data given in Figure~1 of Ref.~\cite{Niyonzima2013} -- namely:

\noindent
~~i) the potential energy curve (PEC) of the ground electronic state of the cation and, 

\noindent
~~~~~for each of the dominant symmetries 
$^2\Sigma^+$, $^2\Pi$ and $^2\Delta$, 

\noindent
~ii) the PECs of the valence dissociative states of the neutral,

\noindent
iii) the quantum defects corresponding to the PECs of the bound Rydberg singly-

\noindent
~~~~~excited states of the neutral,

\noindent
iv) the Rydberg-valence couplings.

Relying on these molecular data, we have performed the inter-nuclear dynamics calculations using the MQDT approach presented in Section~\ref{sec:theory}, for the collisions of low-energy electrons with BeT$^+$ molecular ions. The DR, VE and VdE cross sections have been calculated for all 28 vibrational levels of the target cation. Table~\ref{table:1} shows the energies of the vibrational levels of the cation relative to $v_i^+=0$ as well as its two spectroscopic dissociation energies ($D_ e$ and $D_0$).

The cross sections have been calculated with the inclusion of both direct and indirect mechanisms for the $\Sigma$ and $\Pi$ symmetries, and only the direct mechanism for the $\Delta$ symmetry at the highest (i.e. second) order of theory. The cross sections were calculated for each symmetry for the energy range $10^{-5} - 2.7$ eV of the incident electron, having an energy step of $0.01$ meV. Thus, we managed to cover the energy range up to the dissociation limit of the ion, i.e. 2.704 eV.  The global cross sections - eqs.~(\ref{drxsec1}) and (\ref{vexsec1}) - summed up over all symmetries were averaged over a Maxwellian electron energy distribution in order to obtain the thermal rate coefficients up to a maximum electron temperature of $5000$ K. 

Figures \ref{fig:1}--\ref{fig:5} show the rate coefficients for DR, VE and VdE of BeT$^{+}$ for all the 28 vibrational levels of the electronic ground state. For a given vibrational level of BeT$^+$ the DR and the VdE rate coefficients have mainly monotonically decreasing tendencies as function of the electron temperature, while the VE ones increase. Moreover, the vibrational dependence of the DR rate coefficients show a monotonically decreasing behavior presenting local maxima for $v_i^+=3,8,17,23$ initial target vibrational levels. The DR rate coefficients dominate for $v_i^+\le16$, while for $v_i^+>16$ the monovibrational de-excitations become more important.

The excitations are getting important for higher electron temperatures for each vibrational level of the target. For the lower vibrational levels of the BeT$^+$ cation ($v_i^+\le16$), the VE rates are a notable competitor for  DR or the de-excitation processes above $1000$ K electron temperature  only. For $v_i^+>16$, the competition starts at lower temperatures (even below $300$ K) and, for the highest vibrational levels ($v_i^+\ge24$) and temperatures above 1000 K, VE becomes the dominant process. Notice that the vibrational excitation and de-excitation channels are not necessarily simply connected via detailed balancing, because they involve intermediate states and multi-step processes.

Finally, in order to facilitate kinetic modeling of the beryllium and hydrogen containing plasma and to illustrate the isotopic effect for the beryllium monohydride, we have displayed the rate coefficients for all three isotopologues in the same figure, in two versions.

Figure~\ref{fig:6} shows the ratio of the DR rate coefficients for BeD$^{+}$~\cite{Niyonzima2018} {\it vs} BeH$^{+}$~\cite{Niyonzima2017,Niyonzima2013,Laporta2017}, and for BeT$^{+}$ {\it vs} BeH$^{+}$ molecular cations, as function of the quantum number of the initial vibrational level of the target and of temperature.

In this representation, the isotopic effect increases with the vibrational quantum number of the target, from very weak for v$_i^+$ smaller than $5$ to very significant for v$_i^+$  larger than $8$, being stronger for BeT$^{+}$ than for BeD$^{+}$. This is entirely do to the dependence of the positions of the vibrational levels with respect to the points of crossing between the dissociative PECs and the PEC of the ground state of the target cation.

Figure~\ref{fig:7} shows the DR rate coefficients for BeH$^{+}$, BeD$^{+}$ and BeT$^{+}$ molecular cations as function of the energy of the current vibrational level - relative to the ground ($v_i^+=0$) vibrational level (see table~\ref{table:1}) - for three different electron temperatures, $300$, $1000$ and $5000$ K respectively.  Each point or symbol on these curves belongs to a vibrational level of a different isotopologue. The slight shift in each point is due to the slightly different vibrational level spacing induced by the mass difference of the three isotopologues.

Distinct from the common way of plotting such rate coefficients versus vibrational quantum numbers, which in our case leads to a set of three different curves for each considered electron temperature, we now find that the rate coefficients for the three isotopologues versus vibrational energy instead collapse to a very similar (nearly identical) functional form. With the exception of a few points they are essentially the same. This is a very interesting outcome, similar to the one observed already by Capitelli et al \cite{Capitelli} for pure neutral hydrogen isotopologues. This similarity law provides scalable rate coefficient functions for all isotopologues with reasonable accuracy starting from rate coefficients calculated for one of them only.

In order to allow the versatile implementation of the rate coefficients shown in figuress \ref{fig:1}-\ref{fig:5} in kinetics modeling codes, unlike our previous studies~\cite{Niyonzima2017, Niyonzima2018} where we have used generalised Arrhenius formula, here we considered that the original Arrhenius formula is quite suitable for this (cold) plasma application. We also found that this is falling within the range of accuracy provided by quantum chemistry data.  As in \cite{Little2014} we will use Arrhenius formula as follows:

\begin{equation}\label{eqn:BeH_DR_Interpolation}
k^{fitt}(T) = A \, T^{\alpha}\exp\left[-\frac{B}{T}\right],
\end{equation}

\noindent  for DR and VT (VE and VdE) processes over the electron temperature range $100$~K~$\leq$~$T$~$\leq$~$5000$~K. Assuming a Maxwell-Boltzmann vibrational level distribution we have found that vibrational levels up to $v^+_{max}=10$ and vibrational transition with $\Delta v_{max}=10$ are of importance at $5000$K electron temperatures. The fitting parameters ($A,\alpha,B$) are displayed in Tables \ref{tab:BeT_DR_Interpolation}--\ref{tab:BeT_VT_Interpolation10}. The numerical values, obtained using the equation (\ref{eqn:BeH_DR_Interpolation}), agree with the MQDT-computed ones within a range of errors specified in the caption of each table. A maximal relative deviation and an average Root Mean Square (RMS) were provided per data set.

\section{Conclusions}

The present work completes a large series of studies performed on reactive collisions of the beryllium monohydride cation and its isotopologues with low energy electrons, namely for BeH$^{+}$~\cite{Niyonzima2017,Niyonzima2013,Laporta2017}, for BeD$^{+}$~\cite{Niyonzima2018} and for BeT$^{+}$. Making use of the molecular data set calculated in Ref.~\cite{Roos2009,Niyonzima2013} and by adjusting the nuclear masses, we have performed an MQDT calculation for all 28 vibrational levels of the heaviest beryllium monohydride cation in collisions with electrons having kinetic energy up to the dissociation limit of the ion.

We have provided rate coefficients for dissociative recombination, vibrational excitation and de-excitation of BeT$^{+}$ molecular cation, relevant for detailed kinetic plasma-chemical modeling of the cold edge plasma in fusion devices and for aiding and improving interpretation of spectroscopic beryllium wall release rate experiments. Our present and previously-calculated rate coefficients for the beryllium monohydride isotopologues have been alternatively represented in a comparative way as functions of the target vibrational energy - rather than of the target vibrational quantum number - for different electron temperatures. We found a quite close qualitative and quantitative behavior  for all three of these species. This provides a hitherto apparently unexpected similarity scaling law, significantly reducing complexity of the underlying plasma chemical databases for kinetic plasma chemistry models, to be employed in the above mentioned fusion applications. A quite natural but important extension of this finding, left to future work, would be identification (or disproof) of a similar scaling also for the pure hydrogen molecular ions, which is currently already used in fusion boundary codes merely for simplicity, but so far without a posteriori theoretical confirmation. 

It is known that beryllium and its compounds are toxic, so no experimental data are provided yet. In this regard, our calculations can be considered as reference. The accuracy of the calculated data is based on our method, which was tested and calibrated with other molecular ions where experimental data can be found. However, we are aware that the main source of error comes from the data obtained from quantum chemistry calculations, which provides us potential energy curves and couplings. 

The present article completes the study of {\it low-energy} collisions of electrons with beryllium monohydride cations. Computations of the high-energy cross sections will start soon, relying on the inclusion of further excited dissociative paths and of the vibrational continuum of the ion, this latter element being at the origin of the dissociative excitation. 

Finally we must mention that all the data in tables 1-13 are provided in the Supplementary Materials as ASCII files.  Furthermore, the raw data sets of cross sections and rate coefficients are additionally added. Along the way, the data will be uploaded to the LxCat database \cite{lxc} website.

\begin{acknowledgments}
The authors acknowledge support from F\'ed\'eration de Recherche Fusion par Confinement Magn\'etique (CNRS, CEA and Eurofusion), La R\'egion Normandie, FEDER and LabEx EMC3 via the projects PTOLEMEE, Bioengine, EMoPlaF, COMUE Normandie Universit\'e, the Institute for Energy, Propulsion and Environment (FR-IEPE), the European Union via COST (European Cooperation in Science and Technology) actions TUMIEE (CA17126), MW-Gaia (CA18104) and MD-GAS (CA18212), and ERASMUS-plus conventions between Universit\'e Le Havre Normandie and Politehnica University Timisoara, West University Timisoara and University College London.
We are indebted to Agence Nationale de la Recherche (ANR) via the project MONA, Centre National de la Recherche Scientifique via the GdR TheMS and the DYMCOM project, and the Institute Pascal, University Paris-Saclay for the warm hospitality during the DYMCOM workshop.
NP, FI and JZsM acknowledge support from the UEFISCDI through the mobility project no. PN-III-P1-1.1-MCD-2019-0163.
NP is grateful for the support of the Romanian Ministry of Research and Innovation, project no. 10PFE/16.10.2018, PERFORM-TECH-UPT.
JZsM thanks the financial support of the National Research, Development and Innovation Fund of Hungary, under the FK 19 funding scheme with project no. FK 132989.
The work of DR was carried out under the auspices of the ITER scientists fellow network (ISFN) program, ITER Organization, Route de Vinon-sur-Verdon, CS 90 046 13067 St. Paul-lez-Durance (France).
\end{acknowledgments}

\section*{Appendix}
\noindent{\bf Explanation of figures:}\\ 

\noindent{\bf Figures \ref{fig:1}-\ref{fig:5}}. Dissociative recombination, vibrational excitation and de-excitation  Maxwell rate coefficients for  all the vibrational levels of BeH$^+$ in its ground electronic state.

\begin{tabular}{@{}p{3in}p{5in}@{}}
Ordinate		& Maxwell rate coefficient in cm$^3\cdot$s$^{-1}$ \\
Abscissa	& Electron temperature in K\\
Thick black line	& DR Maxwell rate coefficient \\
Thin coloured lines	& Vibrational excitation rate coefficients\\
Coloured lines	with symbols & Vibrational de-excitation rate coefficients\\
\end{tabular}
\vskip 0.5cm
\noindent{\bf Figure \ref{fig:6}.} Ratios of the dissociative recombination rate coefficients for BeD$^{+}$ {\it vs} BeH$^{+}$ (left),  and for BeT$^{+}$ {\it vs} BeH$^{+}$ (right) molecular cations, as function of the quantum number of the initial vibrational level of the target and of temperature.

\begin{tabular}{@{}p{3in}p{5in}@{}}
Ordinate		& Ratios of dissociative recombination Maxwell rate coefficients \\
Abscissa	& Electron temperature in K\\
\end{tabular}
\vskip 0.5cm

\noindent{\bf Figure \ref{fig:7}}. Dissociative recombination rate coefficient for the three beryllium monohydride isotopologues cations as function of the energy of the initial vibrational level of the target - relative to the ground vibrational level - for three different electron temperatures.

\begin{tabular}{@{}p{3in}p{5in}@{}}
Ordinate		& Dissociative recombination Maxwell rate coefficient in cm$^3\cdot$s$^{-1}$ \\
Abscissa	& Vibrational energy relative to the ground level in eV.\\
Black curves	& DR rate coefficients for BeH$^+$.\\
Blue curves	& DR rate coefficients for BeD$^+$.\\
Red curves	& DR rate coefficients for BeT$^+$.\\
Dashed coloured curves with symbols & DR rate coefficients at $T=300$ K electron temperature.\\
Continuous coloured curves with symbols & DR rate coefficients at $T=1000$ K electron temperature.\\
Dashed-dotted coloured curves with symbols & DR rate coefficients at $T=5000$ K electron temperature.\\
\end{tabular}

\vskip 0.5cm
\noindent In order to facilitate the use of the rate coefficients given in figure \ref{fig:7} we have presented them in tables \ref{tab:isotropic1} and \ref{tab:isotropic2}. The data obtained for each of the three isotopologues are suitable to represent the behavior as a function of the vibrational energy.

\vskip 1.0cm
\noindent{\bf Explanation of tables:}\\ 

\noindent{\bf Table~\ref{tab:BeT_DR_Interpolation}}. List of fitting parameters according to eq.~(\ref{eqn:BeH_DR_Interpolation}), minimum and maximum values of relative difference and root mean squares calculated for dissociative recombination for the $v_i^+=0-10$ vibrational levels of the ground electronic state of BeT$^+$.

\begin{tabular}{@{}p{2.5in}p{6in}@{}}
$v_i^+$		& Initial vibrational level of BeT$^+$ \\[5pt]
Temperature range		& in K\\[5pt]
$A,\,\,\alpha,\,\,B$ & Fitting parameters \\[5pt]
rd$_{max}=\max\limits_{i=1,\dots,n}\left|\frac{k_i(T)-k^{fitt}_i(T)}{k_i(T)}\right|$ & maximum of relative difference\\[7pt]
RMS$=\sqrt{\sum_{i=1}^n\frac{1}{n}\left|\frac{k_i(T)-k^{fitt}_i(T)}{k_i(T)}\right|^2}$ & root mean square
\end{tabular}
\vskip 0.5cm

\noindent{\bf Tables~\ref{tab:BeT_VT_Interpolation0}--\ref{tab:BeT_VT_Interpolation10}}. List of fitting parameters according to eq.~(\ref{eqn:BeH_DR_Interpolation}), minimum and maximum values of relative difference and root mean squares calculated for vibrational transitions ($\Delta v_{max}=10$) for the $v_i^+=0-10$ vibrational levels of the ground electronic state of BeT$^+$.

\begin{tabular}{@{}p{2.5in}p{6in}@{}}
$v_i^+\to v_f^+$		& Vibrational transition of BeT$^+$ \\[5pt]
 $\boldsymbol{v_i^+}$$\to$$\boldsymbol{ v_f^+}$	& Stands for VdE\\[5pt]
Temperature range		& in K\\[5pt]
$A,\,\,\alpha,\,\,B$ & Fitting parameters \\[5pt]
rd$_{max}=\max\limits_{i=1,\dots,n}\left|\frac{k_i(T)-k^{fitt}_i(T)}{k_i(T)}\right|$ & maximum of relative difference\\[7pt]
RMS$=\sqrt{\sum_{i=1}^n\frac{1}{n}\left|\frac{k_i(T)-k^{fitt}_i(T)}{k_i(T)}\right|^2}$ & root mean square
\end{tabular}
\vskip 0.5cm
\noindent{\bf Note}: In Tables~\ref{tab:BeT_DR_Interpolation}--\ref{tab:BeT_VT_Interpolation10} the coefficients were given taking into account a wider temperature range in order to prevent boundary errors. The temperature range was increased with $10\%$ of the boundary values. The global maximal value of relative difference was taken as the maximum among all DR maximum relative differences and the maximum among all VT values for a given target. The average RMS was calculated as the arithmetic mean of the temperature weighted process RMS values. 
\vskip 0.5cm
\noindent{\bf Tables~\ref{tab:isotropic1}-\ref{tab:isotropic2}}.  List of dissociative recombination rate coefficient showing the isotopic effects displayed in figure \ref{fig:7}.

\begin{tabular}{@{}p{2.5in}p{6in}@{}}
Temperature range		& in K\\[5pt]
$v_i^+$		& Initial vibrational level of BeX$^+$ \\[5pt]
$E_{v_i^+} - E_0$	& Energy difference relative to the ground state\\[5pt]
Rate coefficient & in $(cm^3 /s)$\\[5pt]
Data by columns corresponding to isotopologues & $BeX^+\, \in\, \{BeH^+,BeD^+,BeT^+\}$ 
\end{tabular}

\begin{figure}[t]
\centering
\includegraphics[width=0.9\linewidth]{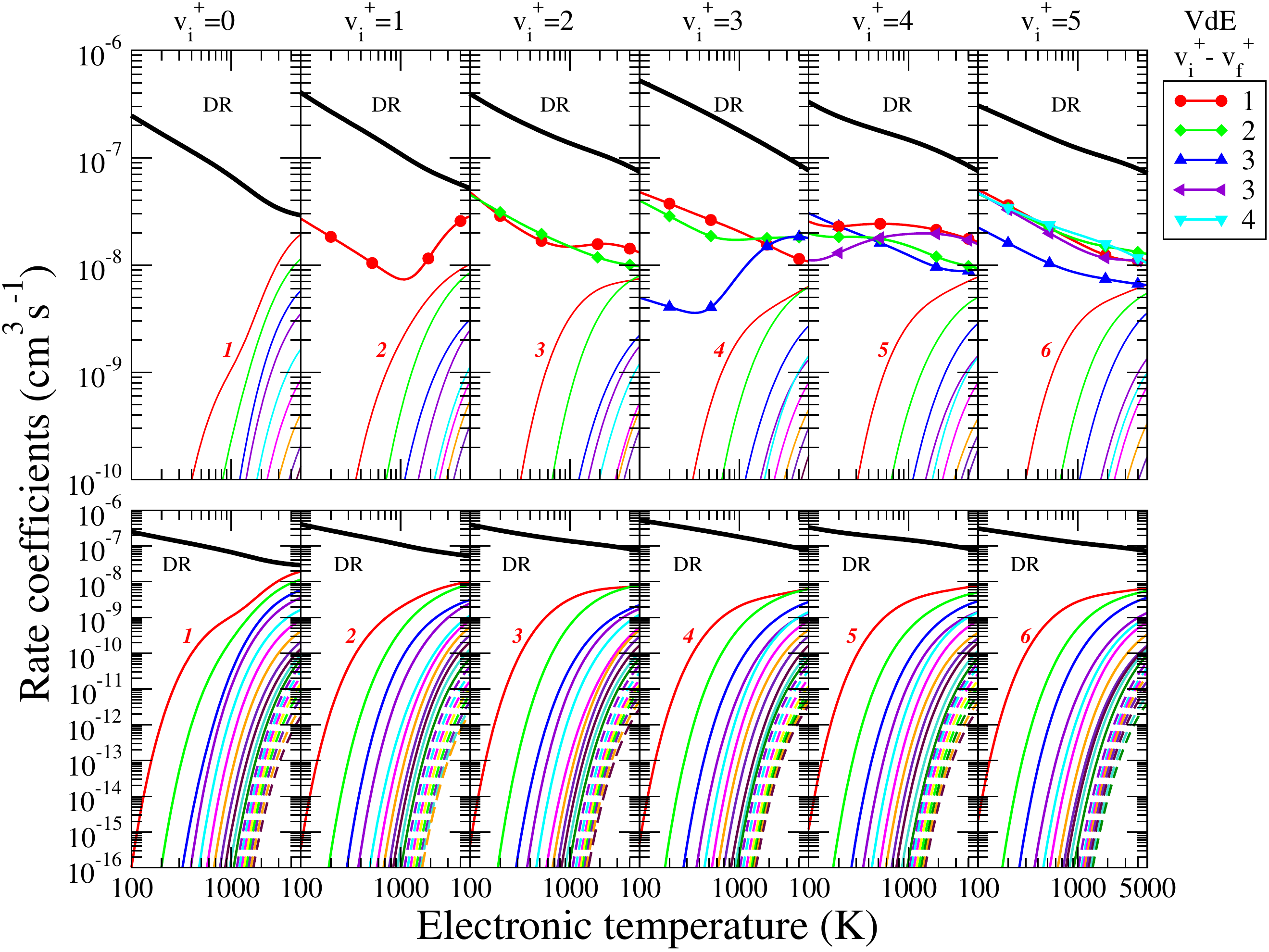}
 \caption{Dissociative recombination (DR, thick black line), vibrational excitation (VE, thin coloured  solid and dashed lines) and vibrational de-excitation (VdE, symbols and thick coloured 	lines) rate coefficients of BeT$^{+}$ in its electronic ground state and for the initial vibrational levels $v_i^+ = 0-5$. Upper panels: for each $v_i^+$ of BeT$^{+}$, all possible de-excitation final vibrational quantum numbers are given, while for the excitation only the first one is labeled. The VE rate coefficients decreases monotonically with the  increase of the final vibrational quantum numbers of the target ion. Lower panels: DR and VE rate coefficients only, with the panels extended down to $10^{-16}$ cm$^{3}\cdot$ s$^{-1}$.}
\label{fig:1}
\end{figure}

 \begin{figure}[t]
\centering
    \includegraphics[width=0.9\linewidth]{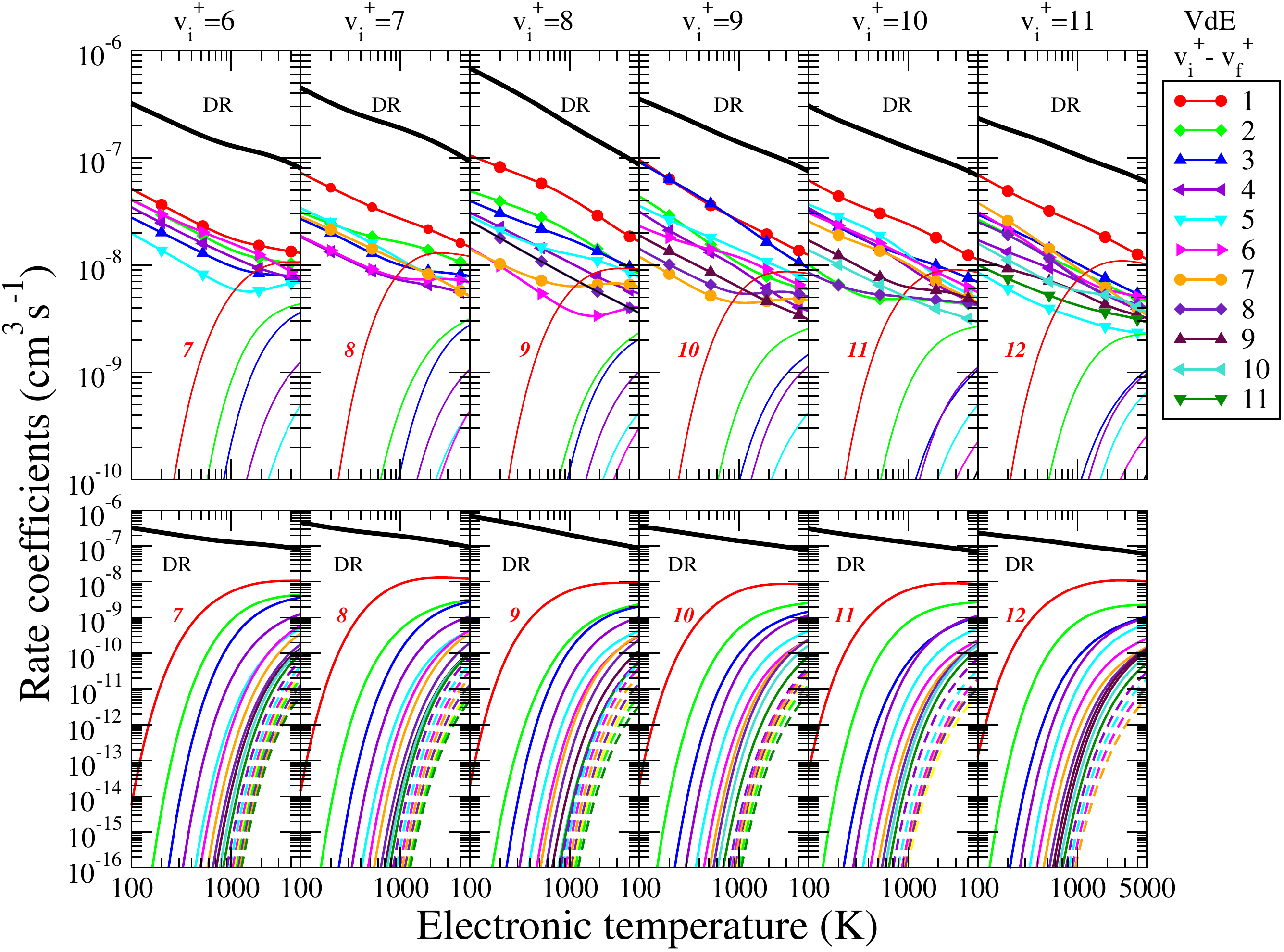}     
     \caption{\label{fig:2} Same as in figure~\ref{fig:1} for $v_i^+ = 6-11$.}
\end{figure}

\clearpage

\begin{figure}[t]
\centering
\includegraphics[width=0.9\linewidth]{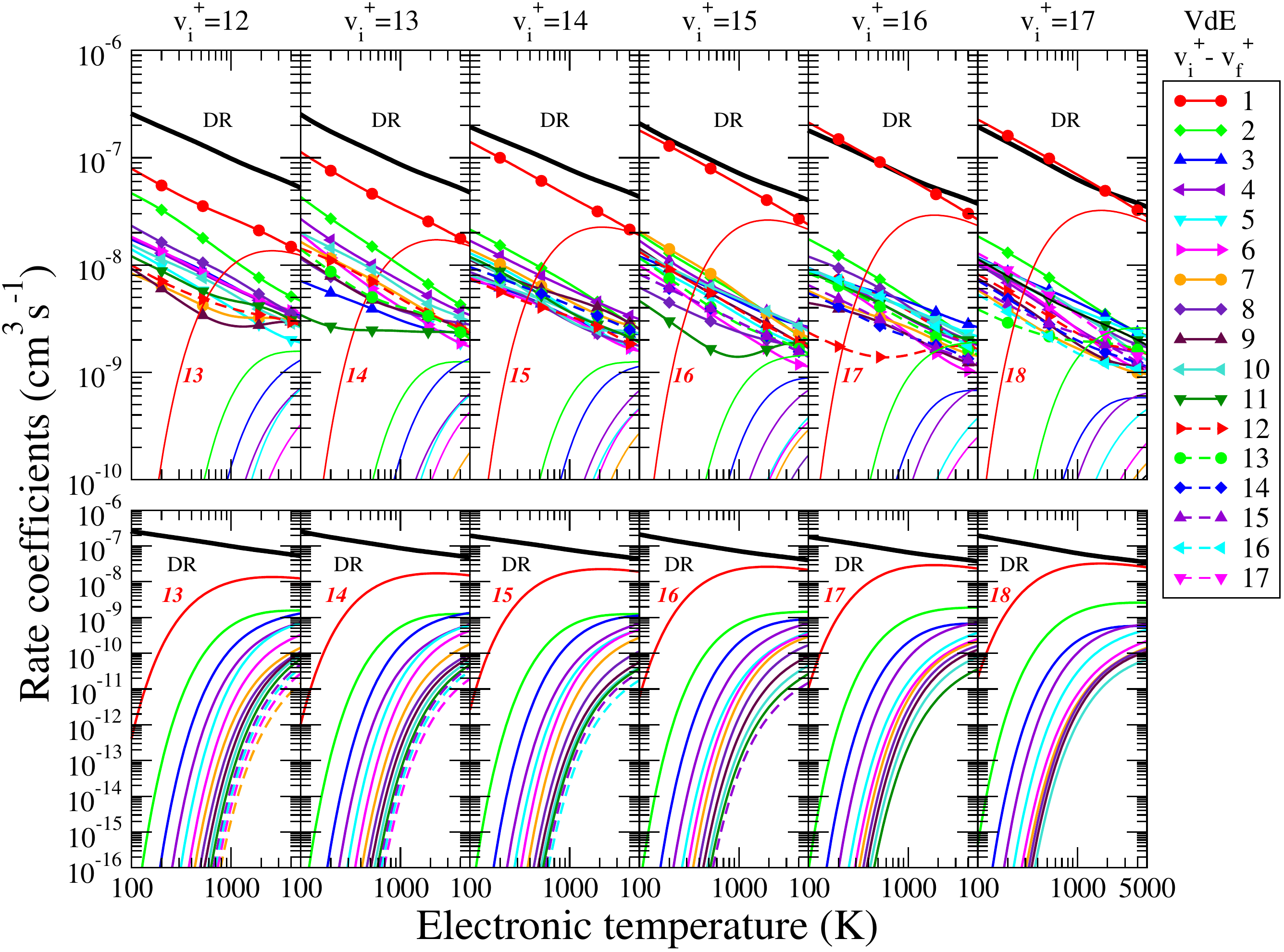} 
\caption{\label{fig:3} Same as in figure~\ref{fig:1} for $v_i^+ = 12-17$.
}
\end{figure}

\begin{figure}[t]
\centering
\includegraphics[width=0.9\linewidth]{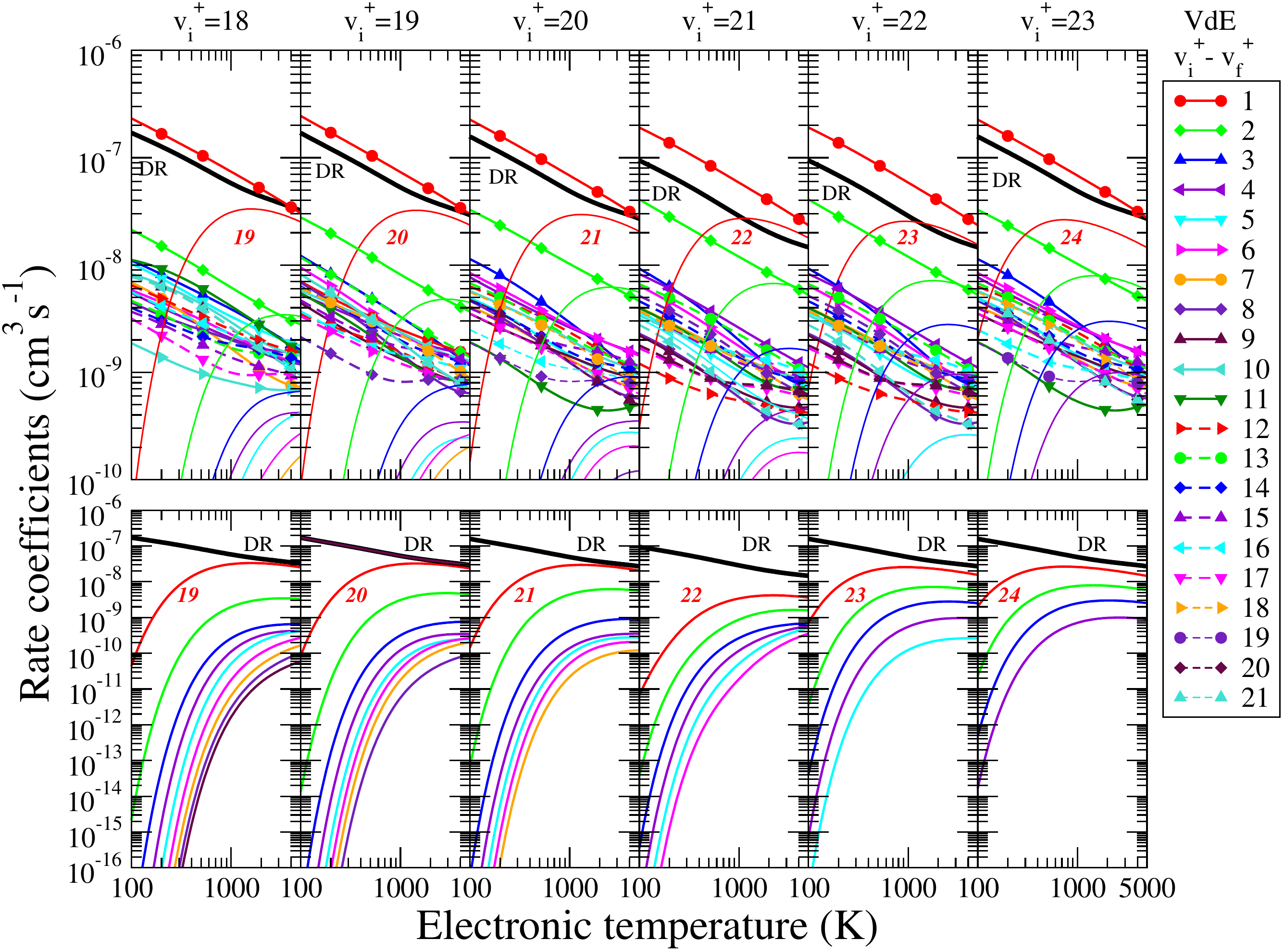} 
\caption{\label{fig:4} Same as in figure~\ref{fig:1} for $v_i^+ = 18-23$.}
\end{figure}

\begin{figure}[t]
\centering
\includegraphics[width=0.75\linewidth]{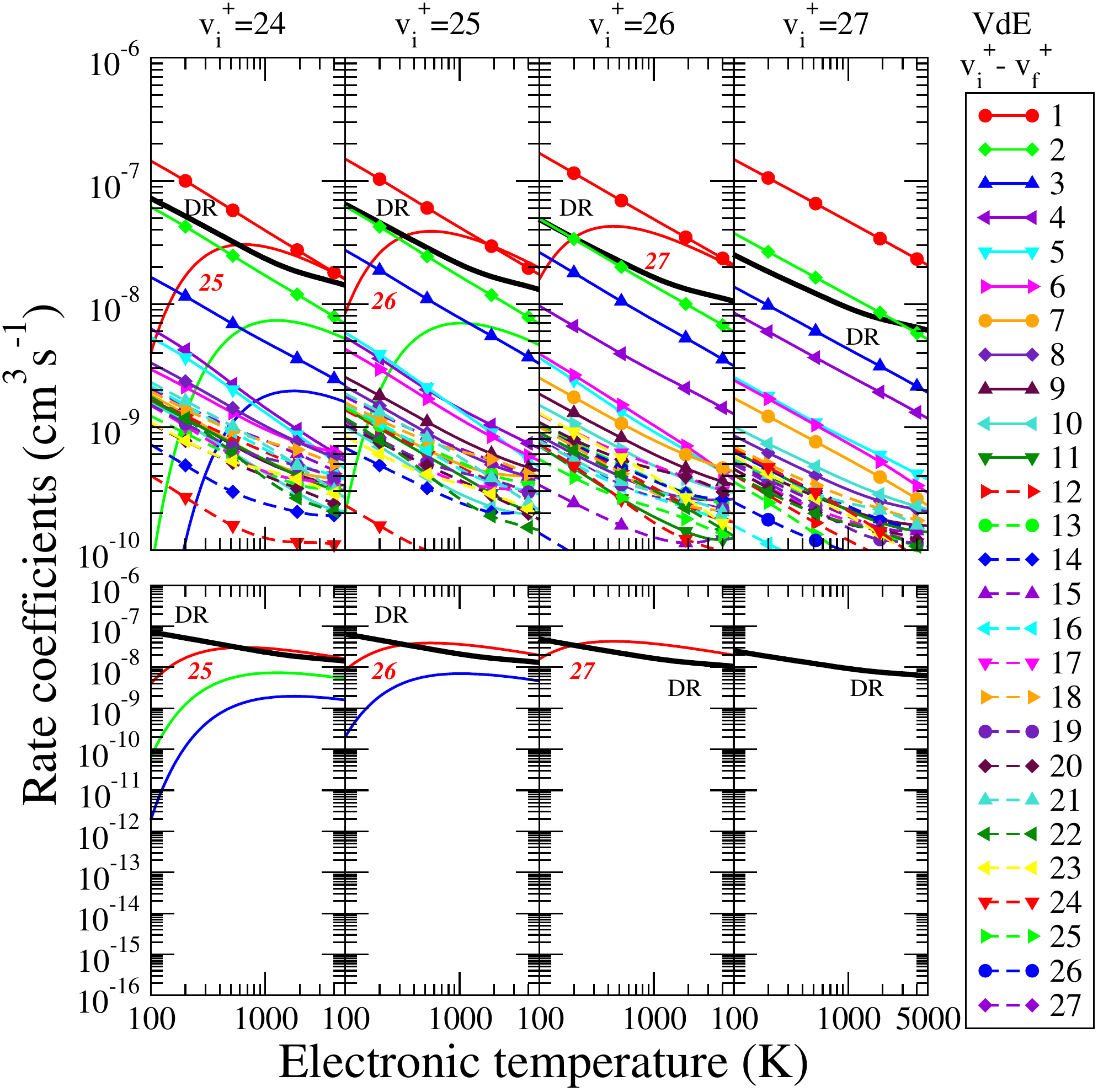}
\caption{\label{fig:5} 
Same as in figure~\ref{fig:1} for $v_i^+ = 24-27$.
}
\end{figure}

 \begin{figure}[t]
\centering
\includegraphics[width=0.9\linewidth]{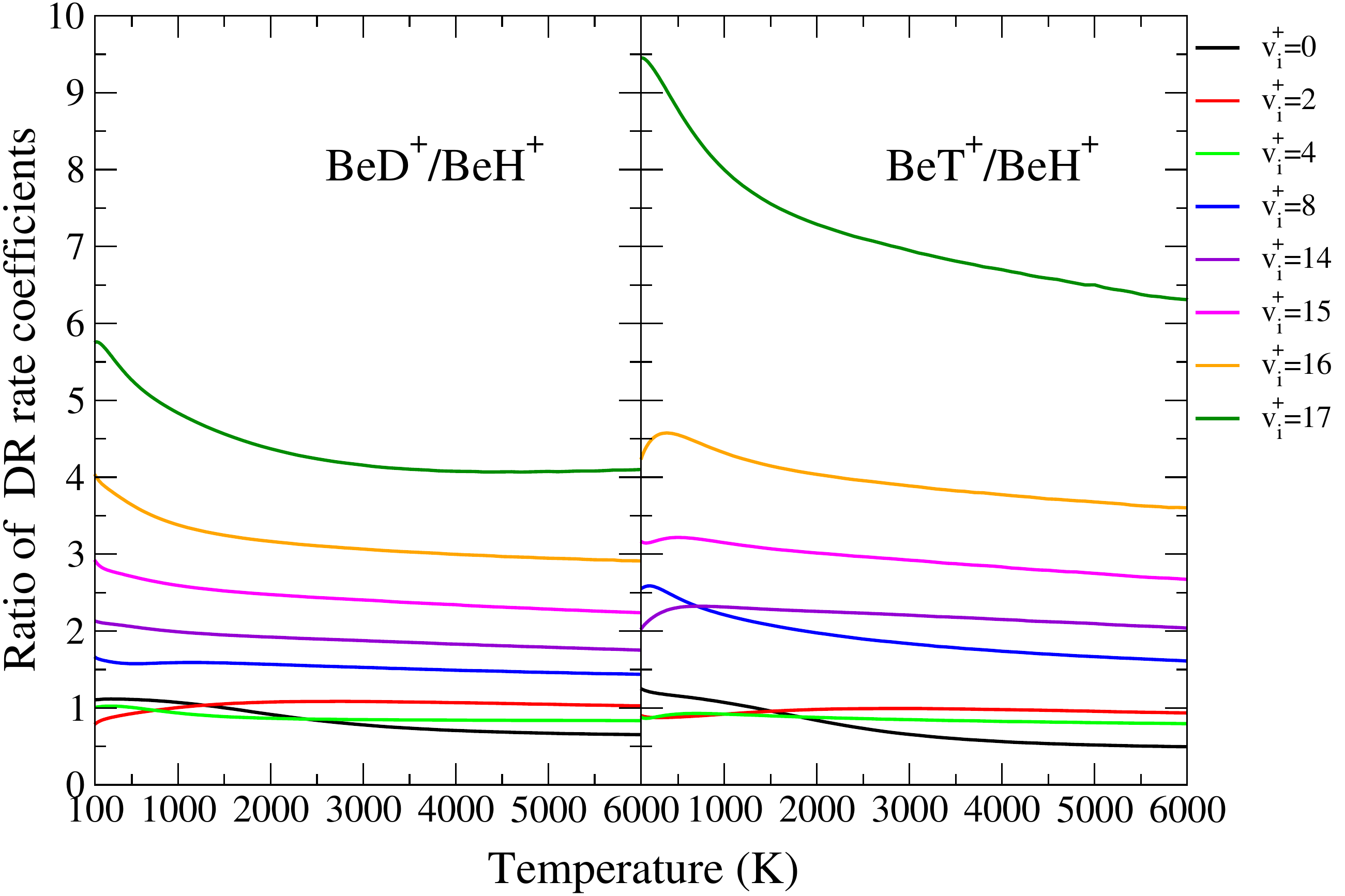}
\caption{\label{fig:6} 
Isotopic effects: Ratios of the dissociative recombination
rate coefficients for 
BeD$^{+}$~\cite{Niyonzima2018} {\it vs} BeH$^{+}$~\cite{Niyonzima2017,Niyonzima2013,Laporta2017} (left),  
and for BeT$^{+}$ {\it vs} BeH$^{+}$ (right) molecular cations, 
as function of the quantum number of the initial vibrational level of the target and of temperature.
}
\end{figure}

 \begin{figure}[t]
\centering
\includegraphics[width=0.9\linewidth]{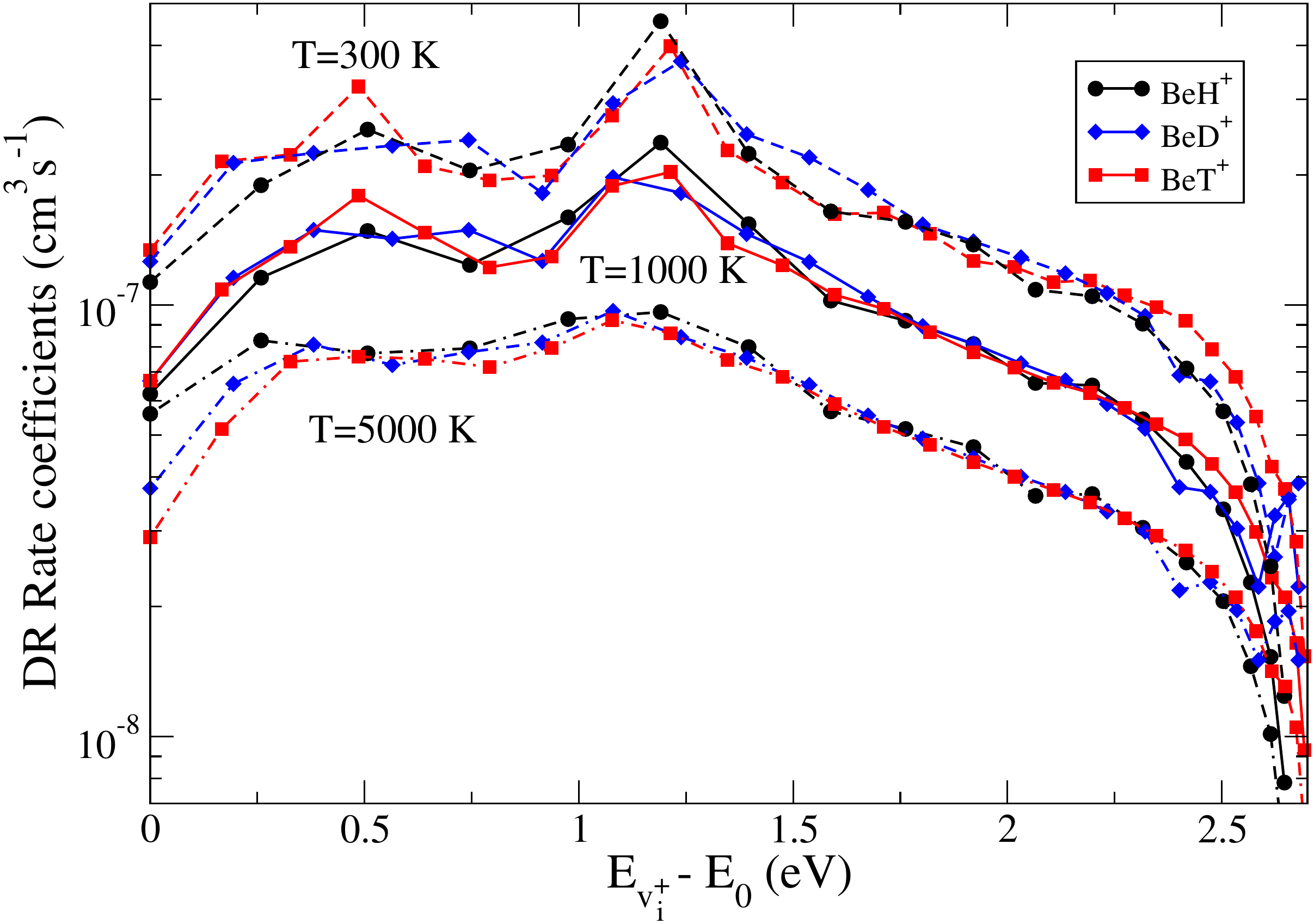}
\caption{\label{fig:7} 
Isotopic effects: Dissociative recombination  rate coefficient of the BeH$^{+}$ (black curves with circles), BeD$^{+}$  
 (blue curves with diamonds) and BeT$^{+}$ (red curves with squares) molecular cations as function of the energy of the initial vibrational level of the target - relative to the ground vibrational level - for three electron temperatures.
} 
\end{figure}

\clearpage

\begin{table}[t]
\caption{Fitting parameters for the DR Maxwell rate coefficients of BeT$^+(v_{i}^{+}=0,\ldots,10)$, for temperatures ranging between $100$ K and $5000$ K as displayed in figures~\ref{fig:1}--\ref{fig:5}. The calculated rate coefficients are reproduced with a maximal relative deviation of $3.242\%$ and with an average Root Mean Square (RMS) of $0.0064$. }
\label{tab:BeT_DR_Interpolation}
\begin{center}
\begin{tabular}{ccccc}
\hline
 $v^+_i$ & Temperature range & A & $\alpha$ & B \\
              & (K) & (cm$^3/$s) & & (K) \\
 \hline
0 & 100 $\le$ T $\le$ 1000 & 3.28197$\cdot$10$^{-6}$ & -0.560182 & 1.28522 \\
  & 1000 $<$ T $\le$ 5000 & 1.2742$\cdot$10$^{-7}$ & -0.193793 & -715.096 \\
1 & 100 $\le$ T $\le$ 1500 & 4.93965$\cdot$10$^{-6}$ & -0.551551 & -3.05762 \\
  & 1500 $<$ T $\le$ 5000 & 5.54357$\cdot$10$^{-7}$ & -0.286297 & -345.175 \\
2 & 100 $\le$ T $\le$ 1500 & 1.59139$\cdot$10$^{-6}$ & -0.359808 & -26.6809 \\
  & 1500 $<$ T $\le$ 5000 & 8.88933$\cdot$10$^{-6}$ & -0.551753 & 429.823 \\
3 & 100 $\le$ T $\le$ 5000 & 6.41436$\cdot$10$^{-6}$ & -0.517293 & 13.1328 \\
4 & 100 $\le$ T $\le$ 1000 & 6.98686$\cdot$10$^{-7}$ & -0.228439 & -30.1691 \\
  & 1000 $<$ T $\le$ 5000 & 8.75990$\cdot$10$^{-6}$ & -0.551726 & 279.519 \\
5 & 100 $\le$ T $\le$ 1000 & 1.81880$\cdot$10$^{-6}$ & -0.393492 & -3.40054 \\
  & 1000 $<$ T $\le$ 5000 & 2.01551$\cdot$10$^{-6}$ & -0.386694 & 140.669 \\ 
6 & 100 $\le$ T $\le$ 5000 & 8.97394$\cdot$10$^{-7}$ & -0.281268 & -28.9724 \\
7 & 100 $\le$ T $\le$ 1000 & 1.08752$\cdot$10$^{-6}$ & -0.258424 & -31.4290 \\
  & 100 $\le$ T $\le$ 5000 & 1.82852$\cdot$10$^{-5}$ & -0.612179 & 350.171 \\
8	&	100	$\le$	T	$\le$	1000	&	8.0781	$\cdot$10$^{-6}$	&	-0.539215	&	0.11067	\\
	&	1000	$<$	T	$\le$	5000	&	1.0082	$\cdot$10$^{-5}$	&	-0.561074	&	11.15700	\\
9	&	100	$\le$	T	$\le$	1000	&	1.7769	$\cdot$10$^{-6}$	&	-0.360498	&	-3.68476	\\
	&	1000	$<$	T	$\le$	5000	&	2.0542	$\cdot$10$^{-6}$	&	-0.388174	&	-3.37710	\\
10	&	100	$\le$	T	$\le$	1000	&	2.7190	$\cdot$10$^{-6}$	&	-0.467593	&	-0.13565	\\
	&	1000	$\le$	T	$\le$	5000	&	2.6040	$\cdot$10$^{-6}$	&	-0.423590	&	128.7260	\\
\hline 
\end{tabular}
\end{center}
\end{table}

\begin{table}[t]
\caption{Fitting parameters for the VT $(\Delta v_{max}=10)$ Maxwell rate coefficients of BeT$^+(v_{i}^{+}=0)$,  for temperatures ranging between $100$ K and $5000$ K, as displayed in figure~\ref{fig:1}. The calculated rate coefficients are reproduced with a maximal relative deviation of $6.073\%$ and with an average RMS of $0.0108$.
}
\label{tab:BeT_VT_Interpolation0}
\begin{center}
\begin{tabular}{ccccc}
\hline
 $v^+_i\rightarrow v^+_f$ & Temperature range & A & $\alpha$ & B \\
              & (K) & (cm$^3/$s) & & (K) \\
\hline
0 $\rightarrow$ 1 & 100 $\le$ T $\le$ 800 & 5.55127$\cdot$10$^{-7}$ & -0.631862 & 1946.6  \\
  & 800 $<$ T $\le$ 1700 & 6.82095$\cdot$10$^{-20}$ & 3.19051 & -1420.11  \\
  & 1700 $<$ T $\le$ 5000 & 2.18465$\cdot$10$^{-6}$ & -0.41325 & 6020.8  \\
0 $\rightarrow$ 2 & 200 $\le$ T $\le$ 540 & 5.61$\cdot$10$^{-8}$ & -0.277382 & 3703.32  \\
  & 540 $<$ T $\le$ 1700 & 1.69648$\cdot$10$^{-14}$ & 1.71428 & 2320.49  \\
  & 1700 $<$ T $\le$ 5000 & 1.8088$\cdot$10$^{-6}$ & -0.439072 & 6608.15  \\
\hline 
\end{tabular}
\end{center}
\end{table}

\begin{table}[t]
\caption{Same as table \ref{tab:BeT_VT_Interpolation0} for $v_i^+= 1$, as displayed in figure \ref{fig:1}, maximal relative deviation of $6.459\%$ and average RMS of $0.0098$.}
\label{tab:BeT_VT_Interpolation1}
\begin{center}
\begin{tabular}{ccccc}
\hline
 $v^+_i\rightarrow v^+_f$ & Temperature range & A & $\alpha$ & B \\
              & (K) & (cm$^3/$s) & & (K) \\
\hline
{\bf 1 $\rightarrow$ 0} & 100 $\le$ T $\le$ 1000 & 4.50249$\cdot$10$^{-7}$ & -0.599873 & 5.30000  \\
  &  1000 $<$ T $\le$ 2500 & 2.43483$\cdot$10$^{-16}$ & 2.21315 & -1823.33  \\
  &  2500 $<$ T $\le$ 5000 & 1.8544$\cdot$10$^{-5}$ & -0.643767 & 4975.26 \\
1 $\rightarrow$ 2 & 100 $\le$ T $\le$ 300 & 2.55835$\cdot$10$^{-7}$ & -0.467812 & 1818.6 \\
  & 300 $<$ T $\le$ 900 & 4.45218$\cdot$10$^{-9}$ & 0.102817 & 1562.25  \\
  & 900 $<$ T $\le$ 5000 & 5.70154$\cdot$10$^{-9}$ & 0.112954 & 1901.61 \\
\hline 
\end{tabular}
\end{center}
\end{table}

\begin{table}[t]
\caption{Same as table \ref{tab:BeT_VT_Interpolation0} for $v_i^+= 2$, as displayed in figure \ref{fig:1}, maximal relative deviation of $5.369\%$ and average RMS of $0.0079$.}
\label{tab:BeT_VT_Interpolation2}
\begin{center}
\begin{tabular}{ccccc}
\hline
 $v^+_i\rightarrow v^+_f$ & Temperature range & A & $\alpha$ & B \\
              & (K) & (cm$^3/$s) & & (K) \\
\hline
{\bf 2 $\rightarrow$ 1} & 100 $\le$ T $\le$ 650 & 1.98340$\cdot$10$^{-7}$ & -0.410144 & -47.2313  \\
  &  650 $<$ T $\le$ 2500 & 2.67382$\cdot$10$^{-9}$ & 0.217131 & -221.517  \\
  &  2500 $<$ T $\le$ 5000 & 1.03204$\cdot$10$^{-5}$ & -0.743861 & 1679.41  \\
{\bf 2 $\rightarrow$ 0} & 100 $\le$ T $\le$ 1000 & 2.65852$\cdot$10$^{-7}$ & -0.422904 & -18.0281  \\
  &  1000 $<$ T $\le$ 5000 & 1.41619$\cdot$10$^{-8}$ & -0.055055 & -438.763  \\
2 $\rightarrow$ 3 & 100 $\le$ T $\le$ 750 & 2.98556$\cdot$10$^{-7}$ & -0.385583 & 1857.56  \\
\hline 
\end{tabular}
\end{center}
\end{table}

\begin{table}[t]
\caption{Same as table \ref{tab:BeT_VT_Interpolation0} for $v_i^+= 3$, as displayed in figure \ref{fig:1}, maximal relative deviation of $4.869\%$ and average RMS of $0.0064$.}
\label{tab:BeT_VT_Interpolation3}
\begin{center}
\begin{tabular}{ccccc}
\hline
 $v^+_i\rightarrow v^+_f$ & Temperature range & A & $\alpha$ & B \\
              & (K) & (cm$^3/$s) & & (K) \\
\hline
{\bf 3 $\rightarrow$ 2} & 100 $\le$ T $\le$ 5000 & 3.46715$\cdot$10$^{-7}$ & -0.410553 & 9.90469  \\
{\bf 3 $\rightarrow$ 1} & 100 $\le$ T $\le$ 650 & 2.12157$\cdot$10$^{-7}$ & -0.394035 & -15.1628  \\
  &  650 $<$ T $\le$ 5000 & 1.58371$\cdot$10$^{-8}$ & 0.015527 & 18.2980  \\
{\bf 3 $\rightarrow$ 0} & 100 $\le$ T $\le$ 380 & 8.75244$\cdot$10$^{-9}$ & -0.163227 & -19.2090  \\
  &  380 $<$ T $\le$ 1000 & 1.36004$\cdot$10$^{-15}$ & 2.15314 & -751.922  \\
  &  1000 $<$ T $\le$ 5000 & 2.54536$\cdot$10$^{-6}$ & -0.531187 & 2127.30  \\
\hline 
\end{tabular}
\end{center}
\end{table}

\begin{table}[t]
\caption{Same as table \ref{tab:BeT_VT_Interpolation0} for $v_i^+= 4$, as displayed in figure \ref{fig:1}, maximal relative deviation of $6.067\%$ and average RMS of $0.0047$.}
\label{tab:BeT_VT_Interpolation4}
\begin{center}
\begin{tabular}{ccccc}
\hline
 $v^+_i\rightarrow v^+_f$ & Temperature range & A & $\alpha$ & B \\
              & (K) & (cm$^3/$s) & & (K) \\
\hline
{\bf 4 $\rightarrow$ 3} & 100 $\le$ T $\le$ 200 & 1.06736$\cdot$10$^{-9}$ & 0.494538 & -89.1826  \\
& 200 $<$ T $\le$ 700 & 3.68736$\cdot$10$^{-8}$ &-0.056086 & 37.2856  \\
&  700 $<$ T $\le$ 5000 & 3.40362$\cdot$10$^{-7}$ & -0.349760 & 255.420  \\
{\bf 4 $\rightarrow$ 2} & 100 $\le$ T $\le$ 400 & 9.47838$\cdot$10$^{-9}$ & 0.0986899 & -26.4064  \\
&  400 $<$ T $\le$ 1000 & 4.68070$\cdot$10$^{-7}$ & -0.4725691 & 170.016  \\
&  1000 $<$ T $\le$ 5000 & 3.33750$\cdot$10$^{-8}$ & -0.157519 & -307.015  \\
\hline 
\end{tabular}
\end{center}
\end{table}

\begin{table}[t]
\caption{Same as table \ref{tab:BeT_VT_Interpolation0} for $v_i^+= 5$, as displayed in figure \ref{fig:1}, maximal relative deviation of $4.841\%$ and average RMS of $0.0047$.}
\label{tab:BeT_VT_Interpolation5}
\begin{center}
\begin{tabular}{ccccc}
\hline
 $v^+_i\rightarrow v^+_f$ & Temperature range & A & $\alpha$ & B \\
              & (K) & (cm$^3/$s) & & (K) \\
\hline
{\bf 5 $\rightarrow$ 4} & 100 $\le$ T $\le$ 1000 & 5.56989$\cdot$10$^{-7}$ & -0.513528 & 3.53340  \\
& 1000 $<$ T $\le$ 2500 & 2.01375$\cdot$10$^{-8}$ &-0.094742 & -436.759  \\
&  2500 $<$ T $\le$ 5000 & 6.94136$\cdot$10$^{-10}$ & 0.295339 & -1236.46  \\
{\bf 5 $\rightarrow$ 3} & 100 $\le$ T $\le$ 1000 & 3.10058$\cdot$10$^{-7}$ &-0.423326 & -4.96023  \\
&  1000 $<$ T $\le$ 5000 & 3.62360$\cdot$10$^{-8}$ & -0.125478 & -116.229 \\
{\bf 5 $\rightarrow$ 2} & 100 $\le$ T $\le$ 1000 & 1.06655$\cdot$10$^{-7}$ &-0.366370& -18.3113  \\
\hline 
\end{tabular}
\end{center}
\end{table}

\begin{table}[t]
\caption{Same as table \ref{tab:BeT_VT_Interpolation0} for $v_i^+= 6$, as displayed in figure \ref{fig:2}, maximal relative deviation of $4.706\%$ and average RMS of $0.0053$.}
\label{tab:BeT_VT_Interpolation6}
\begin{center}
\begin{tabular}{ccccc}
\hline
 $v^+_i\rightarrow v^+_f$ & Temperature range & A & $\alpha$ & B \\
              & (K) & (cm$^3/$s) & & (K) \\
\hline
{\bf 6 $\rightarrow$ 5} & 100 $\le$ T $\le$ 900 & 3.70942$\cdot$10$^{-7}$ & -0.446045 & -7.63342 \\
 & 900 $<$ T $\le$ 5000 & 2.79129$\cdot$10$^{-8}$ & -0.094844 & -201.478  \\
{\bf 6 $\rightarrow$ 4} & 100 $\le$ T $\le$ 1000 & 3.46289$\cdot$10$^{-7}$ & -0.467495 & -2.18171  \\
  &  1000 $<$ T $\le$ 5000 & 1.47926$\cdot$10$^{-8}$ & -0.051965 & -289.979  \\
{\bf 6 $\rightarrow$ 3} & 100 $\le$ T $\le$ 1000 & 2.16284$\cdot$10$^{-7}$ & -0.451211 & -2.10319  \\
  &  1000 $<$ T $\le$ 5000 & 4.54736$\cdot$10$^{-9}$ & 0.054518 & -382.150  \\
\hline 
\end{tabular}
\end{center}
\end{table}

\begin{table}[t]
\caption{Same as table \ref{tab:BeT_VT_Interpolation0} for $v_i^+= 7$, as displayed in figure \ref{fig:2}, maximal relative deviation of $4.507\%$ and average RMS of $0.0049$.}
\label{tab:BeT_VT_Interpolation7}
\begin{center}
\begin{tabular}{ccccc}
\hline
 $v^+_i\rightarrow v^+_f$ & Temperature range & A & $\alpha$ & B \\
              & (K) & (cm$^3/$s) & & (K) \\
\hline
{\bf 7 $\rightarrow$ 6} & 100 $\le$ T $\le$ 5000 & 3.37418$\cdot$10$^{-7}$ & -0.367396 & -16.0055  \\
{\bf 7 $\rightarrow$ 5} & 100 $\le$ T $\le$ 700 & 6.89753$\cdot$10$^{-8}$ & -0.217932 & -21.4795  \\
  &  700 $<$ T $\le$ 1700 & 2.20433$\cdot$10$^{-7}$ & -0.355737 & 150.159  \\
  &  1700 $<$ T $\le$ 5000 & 2.18723$\cdot$10$^{-7}$ & -0.362567 & 50.2353  \\
{\bf 7 $\rightarrow$ 4} & 100 $\le$ T $\le$ 900 & 2.28672$\cdot$10$^{-7}$ & -0.458909 & 2.53416  \\
  &  900 $<$ T $\le$ 5000 & 1.05626$\cdot$10$^{-8}$ & -0.037530 & -200.131  \\
\hline 
\end{tabular}
\end{center}
\end{table}

\begin{table}[t]
\caption{Same as table \ref{tab:BeT_VT_Interpolation0} for $v_i^+= 8$, as displayed in figure \ref{fig:2}, maximal relative deviation of $3.697\%$ and average RMS of $0.0054$.}
\label{tab:BeT_VT_Interpolation8}
\begin{center}
\begin{tabular}{ccccc}
\hline
 $v^+_i\rightarrow v^+_f$ & Temperature range & A & $\alpha$ & B \\
              & (K) & (cm$^3/$s) & & (K) \\
\hline
{\bf 8	\bf $\rightarrow$	7}	&	100 $\le$ T $\le$ 500	&	5.78241$\cdot$10$^{-7}$	&	-0.536410	&	2.54953	\\
			&	500 $<$ T $\le$ 1500	&	6.89449$\cdot$10$^{-6}$	&	-0.712737	&	172.842	\\
			&	1500 $<$ T $\le$ 5000	&	8.52219$\cdot$10$^{-7}$	&	-0.471517	&	-331.533	\\
{\bf 8	{\bf $\rightarrow$}	6}	&	100 $\le$ T $\le$ 500	&	3.16466$\cdot$10$^{-7}$	&	-0.383842	&	10.6386	\\
			&	500 $<$ T $\le$ 1500	&	2.60307$\cdot$10$^{-6}$	&	-0.679387	&	149.899	\\
			&	1500 $<$ T $\le$ 5000	&	2.87412$\cdot$10$^{-7}$	&	-0.426602	&	-396.479	\\
\hline 
\end{tabular}
\end{center}
\end{table}

\begin{table}[t]
\caption{Same as table \ref{tab:BeT_VT_Interpolation0} for $v_i^+= 9$, as displayed in figure \ref{fig:2}, maximal relative deviation of $4.062\%$ and average RMS of $0.0038$.}
\label{tab:BeT_VT_Interpolation9}
\begin{center}
\begin{tabular}{ccccc}
\hline
 $v^+_i\rightarrow v^+_f$ & Temperature range & A & $\alpha$ & B \\
              & (K) & (cm$^3/$s) & & (K) \\
\hline
{\bf 9	{\bf $\rightarrow$}	8}	&	100 $\le$ T $\le$ 500	&	1.44871$\cdot$10$^{-6}$	&	-0.593333	&	-1.62306	\\
			&	500 $<$ T $\le$ 1500	&	3.65915$\cdot$10$^{-7}$	&	-0.392892	&	-69.8892	\\
			&	1500 $<$ T $\le$ 5000	&	2.67847$\cdot$10$^{-7}$	&	-0.367149	&	-264.834	\\
{\bf 9	{\bf $\rightarrow$}	7}	&	100 $\le$ T $\le$ 500	&	1.24170$\cdot$10$^{-6}$	&	-0.695805	&	14.5189	\\
			&	500 $<$ T $\le$ 1500	&	9.79737$\cdot$10$^{-8}$	&	-0.351667	&	-189.513	\\
			&	1500 $<$ T $\le$ 5000	&	1.44986$\cdot$10$^{-8}$	&	-0.117823	&	-505.359	\\
\hline 
\end{tabular}
\end{center}
\end{table}

\begin{table}[t]
\caption{Same as table \ref{tab:BeT_VT_Interpolation0} for $v_i^+= 10$, as displayed in figure \ref{fig:2}, maximal relative deviation of $3.541\%$ and average RMS of $0.0038$.}
\label{tab:BeT_VT_Interpolation10}
\begin{center}
\begin{tabular}{ccccc}
\hline
 $v^+_i\rightarrow v^+_f$ & Temperature range & A & $\alpha$ & B \\
              & (K) & (cm$^3/$s) & & (K) \\
\hline
{\bf 10	{\bf $\rightarrow$}	9}	&	100 $\le$ T $\le$ 500	&	1.92577$\cdot$10$^{-7}$	&	-0.305078	&	-26.8405	\\
			&	500 $<$ T $\le$ 1500	&	9.22604$\cdot$10$^{-7}$	&	-0.512261	&	111.699	\\
			&	1500 $<$ T $\le$ 5000	&	4.18158$\cdot$10$^{-7}$	&	-0.429462	&	-175.701	\\
{\bf 10	{\bf $\rightarrow$}	8}	&	100 $\le$ T $\le$ 650	&	1.10611$\cdot$10$^{-8}$	&	-0.160095	&	-66.0942	\\
			&	650 $<$ T $\le$ 1500	&	6.14458$\cdot$10$^{-9}$	&	-0.031534	&	25.8044	\\
			&	1500 $<$ T $\le$ 5000	&	4.20322$\cdot$10$^{-9}$	&	-0.003797	&	-252.871	\\
\hline 
\end{tabular}
\end{center}
\end{table}

\begin{table}[t]
\caption{Isotopic effects: Dissociative recombination rate coefficient of the BeH$^+$ , BeD$^+$ and BeT$^+$  molecular cations as function of the energy of the initial vibrational level of the target $E_{v_{i}^{+}}-E_{0}$ (relative to the ground vibrational level) for two electron temperatures:  $300$ K and $1000$ K.}
\label{tab:isotropic1}
\begin{center}
\begin{tabular}{c@{\extracolsep{5pt}}ccc@{\extracolsep{5pt}}ccc@{\extracolsep{5pt}}ccc}
\hline
   & \multicolumn{3}{c}{BeH$^{+}$} & \multicolumn{3}{c}{BeD$^{+}$} & \multicolumn{3}{c}{BeT$^{+}$} \\[3pt]
    \cline{2-4} \cline{5-7} \cline{8-10}
 Temperature & $v_{i}^{+}$ & $E_{v_{i}^{+}}-E_{0}$ & Rate coefficient & $v_{i}^{+}$ & $E_{v_{i}^{+}}-E_{0}$ & Rate coefficient & $v_{i}^{+}$ & $E_{v_{i}^{+}}-E_{0}$ & Rate coefficient  \\[3pt]
(K) & & (eV) & (cm$^3/$s) & & (eV) & (cm$^3/$s) & & (eV) & (cm$^3/$s)  \\[3pt]
 \hline
300&0&0.000&1.13043$\cdot$10$^{-7}$&0&0.000&1.26201$\cdot$10$^{-7}$&0&0.000&1.34068$\cdot$10$^{-7}$ \\
 &1&0.259&1.89605$\cdot$10$^{-7}$&1&0.194&2.13453$\cdot$10$^{-7}$&1&0.167&2.15156$\cdot$10$^{-7}$ \\
 &2&0.507&2.55119$\cdot$10$^{-7}$&2&0.381&2.25027$\cdot$10$^{-7}$&2&0.327&2.23157$\cdot$10$^{-7}$ \\
 &3&0.746&2.05097$\cdot$10$^{-7}$&3&0.565&2.34042$\cdot$10$^{-7}$&3&0.486&3.20841$\cdot$10$^{-7}$ \\
 &4&0.975&2.35437$\cdot$10$^{-7}$&4&0.743&2.41174$\cdot$10$^{-7}$&4&0.641&2.09956$\cdot$10$^{-7}$ \\
\hline
\end{tabular}
\end{center}
\end{table}

\begin{table}[t]
\caption{Same as table \ref{tab:isotropic1} for $5000$ K electron temperature.}
\label{tab:isotropic2}
\begin{center}
\begin{tabular}{c@{\extracolsep{5pt}}ccc@{\extracolsep{5pt}}ccc@{\extracolsep{5pt}}ccc}
\hline
   & \multicolumn{3}{c}{BeH$^{+}$} & \multicolumn{3}{c}{BeD$^{+}$} & \multicolumn{3}{c}{BeT$^{+}$} \\[3pt]
    \cline{2-4} \cline{5-7} \cline{8-10}
 Temperature & $v_{i}^{+}$ & $E_{v_{i}^{+}}-E_{0}$ & Rate coefficient & $v_{i}^{+}$ & $E_{v_{i}^{+}}-E_{0}$ & Rate coefficient & $v_{i}^{+}$ & $E_{v_{i}^{+}}-E_{0}$ & Rate coefficient  \\[3pt]
(K) & & (eV) & (cm$^3/$s) & & (eV) & (cm$^3/$s) & & (eV) & (cm$^3/$s)  \\[3pt]
 \hline 
 5000&0&0.000&0.55998$\cdot$10$^{-7}$&0&0.000&0.37608$\cdot$10$^{-7}$&0&0.000&0.29001$\cdot$10$^{-7}$ \\
&1&0.259&0.82794$\cdot$10$^{-7}$&1&0.194&0.65690$\cdot$10$^{-7}$&1&0.167&0.51597$\cdot$10$^{-7}$ \\
 &2&0.507&0.77280$\cdot$10$^{-7}$&2&0.381&0.80984$\cdot$10$^{-7}$&2&0.327&0.73925$\cdot$10$^{-7}$ \\
 &3&0.746&0.79359$\cdot$10$^{-7}$&3&0.565&0.72617$\cdot$10$^{-7}$&3&0.486&0.75943$\cdot$10$^{-7}$ \\
 &4&0.975&0.92771$\cdot$10$^{-7}$&4&0.743&0.77806$\cdot$10$^{-7}$&4&0.641&0.74990$\cdot$10$^{-7}$ \\
\hline
\end{tabular}
\end{center}
\end{table}

\end{document}